# Effect Sizes in Marketing Research: Why Cohen's Local $f^2$ Belongs in the Toolkit

Wolfgang Messner[*]

**Abstract**

In an editorial in the *Journal of Marketing*, Steenkamp et al. (2026) make a valuable and timely intervention by urging marketing scholars to move beyond dichotomous significance testing and to report effect sizes that speak to substantive significance. Their editorial is especially strong in its insistence on exact *p*-values, richer statistical reporting, and closer alignment between rigor and relevance. Yet, their framework omits the local form of Cohen's $f^2$, that is $f_B^2$ as an effect-size measure for the contribution of an individual predictor or predictor block *B* within a multivariable model. That omission matters because much of marketing research relies on regression-type models in which the central theoretical question is not merely whether a model fits globally, but whether a focal construct adds meaningful explanatory power beyond competing predictors and controls. This commentary argues that the $R^2$ foundation of local Cohen's $f_B^2$ is a strength, especially in large-sample settings. Moreover, $f^2$-type local effect sizes can be extended beyond ordinary least squares to multilevel models and, more tentatively, to neural networks and other machine-learning models.

Author affiliation: [*]Darla Moore School of Business, University of South Carolina, Columbia, SC–29208, USA.

Correspondence to email: wolfgang.messner@moore.sc.edu, wolfgang.messner@gmail.com.

Steenkamp et al. (2026) deserve considerable credit for putting empirical execution and substantive significance at the center of the agenda for *Journal of Marketing*. Their editorial rejects the longstanding habit of treating low *p*-values as proxies for importance, calls for exact *p*-values rather than threshold language, and provides a broad table of effect-size metrics spanning association measures, impact measures, and global-fit measures. In a field that has too often conflated statistical significance with practical importance, this is a welcome and necessary recalibration.

However, the effect-size framework they offer is incomplete. For association strength measures, it includes Cohen's *d*, partial eta squared, partial Cohen's $f^2$, odds ratios, Phi, Cramer's *V*, Pearson's *r*, partial correlations, and standardized regression coefficients. For impact measures, it lists marginal effects and elasticities. For global fit, it includes $R^2$. Yet, it does not include Cohen's $f_B^2$ for the local contribution of a single predictor in a regression-type model. That omission is analytically consequential because their framework includes the ANOVA-oriented partial Cohen's $f_p$, but not the regression-oriented Cohen's $f_B^2$, which captures the incremental explanatory contribution of a focal regressor after other variables are already in the model.

This commentary is therefore appreciative in spirit but corrective in substance. The argument is not that Steenkamp et al. (2026) are wrong to emphasize impact metrics such as marginal effects and elasticities, nor that global-fit measures such as $R^2$ should be downgraded. Rather, it is that marketing journals should explicitly add Cohen's $f_B^2$ to their reporting repertoire, because $f_B^2$ answers a distinct and central question in marketing theory testing: How much unique explanatory force does this focal construct add once the rest of the model is already there?

### Why Cohen's $f_B^2$ Fills a Genuine Gap

Selya et al. (2012, p. 1) describe Cohen's $f_B^2$ as a "relatively uncommon, but very informative" measure of local effect size, that is, the effect size of one variable within the context of a multivariate regression model. This motivation is especially important for marketing research: many empirical questions do not ask whether an overall model matters, but whether one theoretically focal variable matters more or less than other simultaneous predictors in the same model. The distinction is between global and local effect sizes, and it is the local question that often drives managerial and theoretical interpretation.

The global form of Cohen's $f^2$ and its local form $f_B^2$ are computed as (Bakeman & McArthur, 1999; Cohen, 1988; Cohen & Cohen, 1983; Selya et al., 2012):

$$f^2 = \frac{R_{AB}^2}{1-R_{AB}^2} \quad \text{and} \quad f_B^2 = \frac{R_{AB}^2 - R_A^2}{1-R_{AB}^2},$$

where $R_{AB}^2$ is the explained variance for the full model including the focal predictor or block $B$ and $R_A^2$ is the explained variance for the model without $B$. The numerator therefore captures the focal variable's incremental contribution beyond the rest of the model $A$, while the denominator scales that increment by the variance left unexplained in the full model. The familiar Cohen benchmarks of .02, .15, and .35 for small, medium, and large effects apply as reference points. Cohen's $f_B^2$ is especially useful because it does not merely describe association, and it does not merely summarize global fit. Instead, it quantifies a predictor's unique explanatory leverage in the model that the researcher actually estimated.

This is a different question from the one answered by standardized beta coefficients, partial correlations, marginal effects, or elasticities. Standardized betas and partial correlations are association measures. Marginal effects and elasticities are impact measures. $R^2$ is a global-fit measure. The own taxonomy of Steenkamp et al. (2026) therefore creates a conceptual opening for $f_B^2$: it is a local, variance-accounted-for effect size that complements rather than duplicates the metrics already in their framework. In other words, $f_B^2$ belongs in the toolkit not because it replaces existing measures, but because it captures something they do not.

That distinction is especially important in marketing because much of the field's nonexperimental work relies on multivariable regression-type models in which theoretically adjacent constructs compete for explanatory space. In such settings, a focal predictor can produce a low $p$-value simply because sample size is large, yet add very little incremental explained variance once controls and neighboring constructs are included. Conversely, a predictor may have modest raw magnitude but still make a meaningful unique contribution in a crowded model. Cohen's $f_B^2$ is designed to discipline such interpretations. It asks whether the focal construct truly earns its role in the nomological network, rather than merely surviving threshold-based hypothesis testing.

### The $R^2$ Foundation of $f_B^2$

One reason Cohen's $f_B^2$ deserves more attention is that it is anchored in $R^2$. That feature gives it a direct variance-accounted-for interpretation that is intuitive to readers across methodological traditions (Dedecker et al., 2025). Put simply, $R^2$ captures the share of variance in the outcome explained by the model, and $f_B^2$ converts the incremental contribution of the focal predictor $B$ into a standardized local effect size. On that basis, $f_B^2$ has a natural interpretive clarity: it is a local transformation of the same variance-accounted-for logic, but applied to one predictor's incremental contribution (for an application, see e.g. De La Rosa et al., 2025; Messner et al., 2025).

To be sure, sample $R^2$ is not flawless in finite samples. The literature has long recognized the need for sample-size planning and bias correction (Kelley, 2008; Olkin & Pratt, 1958). If the objection to Cohen's $f^2$ is that it depends on $R^2$, the appropriate response is that $R^2$ can indeed be optimistic in smaller samples, and that is exactly why adjusted $R^2$, shrinkage estimators, and interval estimation exist. This is a manageable statistical issue, not a conceptual defect.

The large-sample case, which is often the more relevant one in contemporary big-data analytics, is even stronger (Khalilzadeh & Tasci, 2017). Since the treatment by Fisher (1928), the sampling distribution of the multiple correlation coefficient has been studied extensively (e.g., Dedecker et al., 2025; Lee, 1971; Ogasawara, 2006; Soper, 1929). Taken together, this literature suggests that, in adequately powered large samples, $R^2$ is not a fragile base for local effect-size reasoning, but one of the better-understood quantities in the regression toolkit.

By implication, the same large-sample stability extends to $f^2$. Because local $f_B^2$ is a transformation of reduced-model $R_A^2$ and full-model $R_{AB}^2$ estimates, the stability of those inputs carries over to the derived effect size so long as the denominator $1 - R_{AB}^2$ is not vanishingly small. That is not a heroic assumption in most marketing settings, where even useful models rarely approach perfect fit. In the large samples common to panel, customer management, online platform, scanner, and digital-trace data, the $R^2$ foundation of $f_B^2$ is therefore an advantage because it yields a dimensionless metric whose statistical behavior becomes more stable, not less, as sample information accumulates.

## Beyond OLS Regression

Cohen's $f_B^2$ is not confined to OLS regression. Selya et al. (2012) and Lorah (2018) discuss how to compute local $f_B^2$ in mixed-effects and multilevel models. Their contribution is important because marketing data are frequently longitudinal, repeated, nested, or otherwise hierarchical. In other words, the methodological frontier has already moved beyond the idea that local effect size is only an OLS concern. It travels well across common model classes whenever researchers can define a meaningful and defensible explained-variance decomposition for the focal effect (for an application, see e.g. Effron et al., 2018).

The case for explicitly adding $f_B^2$ to the toolkit becomes even stronger once we look to machine learning. Messner (2023) argues that one of the major barriers to using deep neural networks in research models is that they lack the familiar inferential outputs that social-science researchers expect: direction, significance, and effect size. He proposes a model-agnostic framework for scalar regression problems that estimates effect sizes akin to Cohen's $f_B^2$ for the input variables of a deep artificial neural network. The framework uses a permutation-based logic: compute baseline model performance, permute one predictor at a time, translate the resulting performance loss into an $R^2$-like quantity, and then derive a local $f_B^2$-style measure from the change in explained variance. Cohen's $f^2$ is informative because it can be localized to the contribution of each variable within a fuller model (for an application, see e.g. Messner, 2024).

The broader implication is that $f^2$ is not merely a legacy statistic tied to linear models of the twentieth century. Rather, it is becoming a bridge concept between interpretable statistical modeling and contemporary machine-learning workflows. The machine-learning extension is approximate and probabilistic rather than deterministic, but it demonstrates that the conceptual logic of local, $R^2$-based effect sizes is sufficiently robust to travel beyond conventional linear models.

## Conclusion: What Marketing Researchers Should Do

When marketing researchers make focal claims about the importance of individual predictors in linear, generalized, or hierarchical models, they should report: (1) exact *p*-values, (2) interval estimates of the coefficients, and (3) a local effect-size measure such as $f_B^2$ when a meaningful $R^2$-based decomposition is possible. For smaller samples, researchers should consider adjusted or corrected $R^2$ estimates or confidence intervals around $R^2$-based quantities.

Global $R^2$ remains indispensable for model-level fit. But local $f_B^2$ adds something distinct: a standardized answer to whether the focal variable contributes materially beyond what the rest of the model already explains. As marketing research increasingly prizes substantive significance, that is not a marginal question but a central one.

Steenkamp et al. (2026) have done the field a service by insisting that methodological rigor must serve substantive relevance and that effect-size reporting must complement, rather than trail behind, significance testing. The argument of this commentary is that their framework can be made stronger by adding Cohen's $f_B^2$ to the effect-size toolkit. This statistic most directly addresses local explanatory importance in the regression-based models that dominate much of marketing research. Its $R^2$ foundation is statistically mature, especially in large samples, and its logic extends beyond OLS to multilevel modeling and machine-learning applications.